\documentclass[preprint,prc,superscriptaddress,unsortedaddress,showpacs,preprintnumbers,amsmath,amssymb]{revtex4}
\usepackage{tipa}

\usepackage{graphicx}
\usepackage{dcolumn}
\usepackage{bm}

\def\beq{\begin{equation}}
\def\eeq{\end{equation}}
\def\bea{\begin{eqnarray}}
\def\eea{\end{eqnarray}}

\def\fun#1#2{\lower3.6pt\vbox{\baselineskip0pt\lineskip.9pt
  \ialign{$\mathsurround=0pt#1\hfil##\hfil$\crcr#2\crcr\sim\crcr}}}

\preprint{}

\begin{document}

\title{FFLO state with angle-dependent gap in Asymmetric Nuclear Matter}

\author{Xin-le Shang}\email[ ]{shangxinle@impcas.ac.cn}
 \affiliation{Institute
of Modern Physics, Chinese Academy of Sciences, Lanzhou 730000,
China}\affiliation{State Key Laboratory of Theoretical Physics,
Institute of Theoretical Physics, Chinese Academy of Sciences,
Beijing 100190, China}

\author{Pei Wang}
\affiliation{Institute of Modern Physics, Chinese Academy of
Sciences, Lanzhou 730000, China} \affiliation{University of
Chinese Academy of Sciences, Beijing, 100049, People's Republic of
China}

\author{Peng Yin}
\affiliation{Institute of Modern Physics, Chinese Academy of
Sciences, Lanzhou 730000, China} \affiliation{University of
Chinese Academy of Sciences, Beijing, 100049, People's Republic of
China}
\author{Wei Zuo}
\affiliation{Institute of Modern Physics, Chinese Academy of
Sciences, Lanzhou 730000, China}\affiliation{State Key Laboratory of Theoretical Physics, Institute of Theoretical Physics, Chinese Academy of
Sciences, Beijing 100190, China}

\begin{abstract}
We consider the FFLO state with angle-dependent gap (ADG) for
arbitrary angle $\theta_{0}$ between the direction of the Cooper
pair momentum and the symmetry-axis of ADG in asymmetric nuclear
matter. We find two kinds of locally stable states, i.e., the
FFLO-ADG-Orthogonal and FFLO-ADG-Parallel states, which correspond
to $\theta_{0}=\frac{\pi}{2}$ and $\theta_{0}=0$, respectively.
Furthermore, the FFLO-ADG-Orthogonal state is located at low
asymmetry, whereas the FFLO-ADG-Parallel state is favored for
large asymmetry. The critical isospin asymmetry $\alpha_{c}$,
where the superfluid vanishes, is enhanced largely by considering
the Cooper pair momentum with ADG.

\end{abstract}

\maketitle

\section{Introduction}
Neutron-proton (n-p) pair correlations are potentially important
in a number of physical contexts, including the mechanism of the
deuteron formation in heavy-ion collisions\cite{Snm2} at
intermediate energies and supernova\cite{Sn1,Sn2,Sn3}. In the
context of nuclear structure, as evidenced by the recent
experimental findings\cite{Exp} on excited states in $^{92}$Pd,
heavy nuclei may feature spin-aligned n-p pairs, and moreover the
exotic nuclei with extended halos provide a locus for n-p pairing.
In addition, the n-p pairing may play a major role in determining
the cooling and rotation dynamics in the model of ``nucleon stars"
which permits pion or kaon condensation\cite{Ns}.

The occurrence of the n-p pairing crucially depends on the overlap
between the neutron and proton Fermi surfaces. The pairing
correlation is suppressed when the system is driven out of the
isospin-symmetric case. At low temperature, the thermal smearing
of the Fermi surfaces promotes the pairing, but, is ineffective
when the separation between the two Fermi surfaces is large
compared to the temperature. However, a shift of the neutron and
proton Fermi spheres with respect to each other, which results in
a nonzero total momentum of the Cooper pairs, is expected to
enhance the overlap between the two Fermi surfaces. The overlap
region then provides the kinematical phase space for pairing
phenomena to occur. In this configuration space such a condensate
forms a periodic lattice with finite shear modulus. The resulting
inhomogeneous superconducting state is called FFLO
state\cite{ff,lo}. Another possible mechanism enhancing the
overlap between the separated Fermi surfaces of neutron and proton
in asymmetric nuclear matter is the deformation of the two Fermi
surfaces\cite{dfs,dfsn}, which causes the formation of DFS
(deformed Fermi surfaces) state. Both these two kinds of
configurations imply an anisotropic quasiparticle spectrum. On the
other hand, the previous studies of the DFS state\cite{dfsn} and
the FFLO sate\cite{ffn} adopt the angle-averaging procedure which
has been proved to be a quite good approximation in symmetry
nuclear matter\cite{aap} by considering the pairing gap as an
isotropic one. In fact, our previous work\cite{min} indicates the
angle dependence of the pairing gap should be taken into account
when calculating the pairing gap in asymmetric nuclear matter at
low temperature. In Ref.\cite{min}, we have proposed an
axi-symmetric angle-dependent gap (ADG) state which corresponds to
the axi-symmetric deformation of the neutron and proton Fermi
spheres. In the ADG state, the rotational symmetry is broken
spontaneously
and there exists a symmetry breaking axis. While in the FFLO
configuration, both the rotational and translational symmetries
are spontaneously broken 
and the axis of symmetry breaking is along the direction of the
total Cooper pair momentum. To determinate the structure of the
true ground state for bulk isospin-asymmetric nuclear matter, we
should consider the FFLO and ADG state together within the same
model as done in Ref.\cite{ffn,min}.

The purpose of the present paper is a combined treatment of the
FFLO and the axi-symmetric angle-dependent gap (ADG) phases within
the same model as in Ref.\cite{ffn,min} for asymmetric nuclear
matter. Arbitrary angles
between the two symmetry breaking axes are considered to find out
the favored angle for the ground state. In the calculations we
only take into account the $^{3}SD_{1}$ partial-wave channel,
which dominates the pairing interaction at low
density\cite{Snm1,Sdd,Sdd2,Sdd3,Sdd4,Sdd5}. The paper is organized
as follows: In Sec. II we derive the gap equations within the
Gorkov formalism, which include the effects of the finite momentum
of the Cooper pairs and the axi-symmetric angle dependence of
pairing gap. The numerical solutions of these equations are shown
in Sec. III, where we discuss the phase diagram of the combined
FFLO and ADG states and their properties for different angles
between
two symmetry breaking axes at finite temperature. We finish in
Sec. IV with a summary of our results and present our conclusions.

\section{Formalism}
At low densities, the isospin singlet $^{3}SD_{1}$ pairing channel
dominates the attractive pairing force. In this case we can
consider $^{3}SD_{1}$ channel alone, and the gap function can be
expanded\cite{aap,tt} according to
\begin{eqnarray}
\Delta_{\sigma_{1},\sigma_{2}}(\textbf{k})
=\sum_{l,m_{j}}\Delta_{l}^{m_{j}}(k)[G_{l}^{m_{j}}(\textbf{\^{k}})]_{\sigma_{1},\sigma_{2}},
\end{eqnarray}
with the elements of the spin-angle matrices
\begin{eqnarray}
[G_{l}^{m_{j}}(\textbf{\^{k}})]_{\sigma_{1},\sigma_{2}}\equiv
\langle\frac{1}{2}\sigma_{1},\frac{1}{2}\sigma_{2}\mid1\sigma_{1}+\sigma_{2}\rangle
\langle1\sigma_{1}+\sigma_{2},lm_{l}\mid1m_{j}\rangle Y_{l}^{m_{l}}(\textbf{\^{k}}),\nonumber\\
\end{eqnarray}
where $m_{j}$ and $m_{l}$ are the projections of the total angular
momentum $j=1$ and the orbit angular momentum $l=0,2$ of the pair,
respectively. The $Y_{l}^{m_{l}}(\textbf{\^{k}})$ denotes the
spherical harmonic with $\textbf{\^{k}}\equiv\textbf{k}/k$. The
anomalous density matrix follows the same expansion. Moreover the
time-reversal invariance implies,
\begin{eqnarray}
\Delta_{\sigma_{1}, \sigma_{2}}(\textbf{k})
=(-1)^{1+\sigma_{1}+\sigma_{2}}\Delta_{-\sigma_{1},-\sigma_{2}}^{*}(\textbf{k}).
\end{eqnarray}
Accordingly, the pairing gap matrix $\Delta(\textbf{k})$ in spin
space possesses the property
\begin{eqnarray}
\Delta(\textbf{k})\Delta^{\dag}(\textbf{k})=ID^{2}(\textbf{k}),
\end{eqnarray}
i.e., the gap function has the structure of a ``unitary triplet"
state \cite{aap}. $I$ is the identity matrix and $D(\textbf{k})$
is a scalar quantity in spin space.
\subsection{The quasiparticle spectrum}
Once the isospin singlet $^{3}SD_{1}$ channel has been selected,
the pairing gap is an isoscalar and the isospin indices can be
dropped out. The proton and neutron propagators following from the
solution of the Gorkov equations are ($\hbar=1$)
\begin{eqnarray}
\textbf{G}_{\sigma,\sigma^{'}}^{(p/n)}(\textbf{k},\omega_{m})=
-\delta_{\sigma,\sigma^{'}}\frac{\imath\omega_{m}+\xi_{\textbf{k}}\mp
\delta\varepsilon_{\textbf{k}}}
{(\imath\omega_{m}+E_{\textbf{k}}^{+})(\imath\omega_{m}-E_{\textbf{k}}^{-})},
\end{eqnarray}
where $\omega_{m}$ are the Matsubara frequencies. In the right
hand side of Eq.(5), the negative and positive signs
correspond to proton and  neutron, respectively. The
neutron-proton anomalous propagator matrix in spin space has the
form
\begin{eqnarray}
\textbf{F}^{\dag}(\textbf{k},\omega_{m})=-\frac{\Delta^{\dag}(\textbf{k})}
{(\imath\omega_{m}+E_{\textbf{k}}^{+})(\imath\omega_{m}-E_{\textbf{k}}^{-})}.
\end{eqnarray}
The quasiparticle excitation spectra are determined by finding the
poles of the propagators in the Gorkov equations,
\begin{eqnarray}
E_{\textbf{k}}^{\pm}=\sqrt{\xi_{\textbf{k}}^{2}+\frac{1}{2}Tr(\Delta\Delta^{\dag})\pm\frac{1}{2}\sqrt{[Tr(\Delta\Delta^{\dag})]^{2}-
4\det(\Delta\Delta^{\dag})}}\pm \delta\varepsilon_{\textbf{k}},
\end{eqnarray}
where
$\xi_{\textbf{k}}=\frac{1}{2}(\varepsilon_{\textbf{k}}^{p}+\varepsilon_{\textbf{k}}^{n})$,
and
$\delta\varepsilon_{\textbf{k}}=\frac{1}{2}(\varepsilon_{\textbf{k}}^{p}-\varepsilon_{\textbf{k}}^{n})$.
The single-particle energy spectra (in this paper we consider the
free single particle energy spectrum) of neutron and proton are
given by
\begin{eqnarray*}
\varepsilon_{\textbf{k}}^{(n)}=\frac{(\textbf{Q}+
\textbf{k})^{2}}{2m}-\mu^{(n)},\varepsilon_{\textbf{k}}^{(p)}=\frac{(\textbf{Q}-
\textbf{k})^{2}}{2m}-\mu^{(p)},
\end{eqnarray*}
with the chemical potential for neutrons and protons $\mu^{(n/p)}$
which are derived selfconsistently from the BCS theory. The
neutron and proton Fermi spheres are shifted with respect to each
other and the Cooper pairs have a total pair momentum
$2\textbf{Q}$. The symmetric and asymmetric parts of the spectrum
(which are even and odd with respect to the time-reversal
symmetry) are performed as
\begin{eqnarray}
\xi_{\textbf{k}}\equiv\frac{\textbf{k}^{2}}{2m}+\frac{\textbf{Q}^{2}}{2m}-\mu,
& &
\delta\varepsilon_{\textbf{k}}\equiv\delta\mu-\frac{\textbf{k}\cdot\textbf{Q}}{2m}.
\end{eqnarray}
Here $\mu=(\mu^{(n)}+\mu^{(p)})/2$ and
$\delta\mu=(\mu^{(n)}-\mu^{(p)})/2$ are the average and relative
chemical potentials, respectively. The
$\frac{\textbf{k}\cdot\textbf{Q}}{2m}$ term due to the Cooper pair
momentum reduces the suppression from the mismatched Fermi surface
$\delta\mu$ in certain directions. Using the ``unitary" property
Eq. (4), the quasiparticle spectra can be simplified as follows
\begin{eqnarray}
E_{\textbf{k}}^{\pm}\equiv
E^{\pm}(\textbf{k},\textbf{Q})=\sqrt{\xi_{\textbf{k}}^{2}+D^{2}(\textbf{k})}\pm
\delta\varepsilon_{\textbf{k}}.
\end{eqnarray}
The limit $\delta\varepsilon_{\textbf{k}}\rightarrow0$ corresponds
to the BCS pairing in symmetric nuclear matter, whereas in
asymmetric nuclear matter the spectra Eq.(9) split into two
branches due to the isospin asymmetry ($\delta\mu\neq0$) and the
finite-momentum of the Cooper pair ($\textbf{Q}\neq0$).

\subsection{The FFLO-ADG gap equations}

In the present ``unitary triplet" case, the gap equation at finite
temperature can be written in the standard form
\begin{eqnarray}
\Delta_{\sigma_{1},
\sigma_{2}}(\textbf{k},\textbf{Q})=-\sum_{\textbf{k}^{'}}\sum_{\sigma_{1}^{'},
\sigma_{2}^{'}}
<\textbf{k}\sigma_{1},-\textbf{k}\sigma_{2}\mid V\mid\textbf{k}^{'}\sigma_{1}^{'},-\textbf{k}^{'}\sigma_{2}^{'}>\nonumber\\
\times\frac{\Delta_{\sigma_{1}^{'},
\sigma_{2}^{'}}(\textbf{k}^{'},\textbf{Q})}{2\sqrt{\xi_{\textbf{k}^{'}}^{2}+D^{2}(\textbf{k}^{'})}}
[1-f(E_{\textbf{k}^{'}}^{+})-f(E_{\textbf{k}^{'}}^{-})],
\end{eqnarray}
where $f(E)=[1+\exp(\beta E)]^{-1}$ is the Fermi distribution
function and $V$ is the interaction in the $^{3}SD_{1}$ channel.
$\beta^{-1}=k_{B}T$, where $k_{B}$ is the Boltzmann constant and
$T$ is the temperature. It is worth noting that the orientation of
$\textbf{Q}$ only affects the value of
$\delta\varepsilon_{\textbf{k}}$ in Eq.(9) through the angle
between $\textbf{Q}$ and $\textbf{k}^{'}$. Using the properties of
spherical harmonics, we can express
$\delta\varepsilon_{\textbf{k}}$ as
\begin{eqnarray}
\delta\varepsilon_{\textbf{k}}&&=\delta\mu-\frac{\textbf{k}^{'}\cdot\textbf{Q}}{2m}=\delta\mu-\frac{k^{'}Q}{2m}\cos(\widehat{\textbf{k}^{'}\textbf{Q}})\nonumber\\
&&=\delta\mu-\frac{k^{'}Q}{2m}[\sin\theta_{0}\sin\theta\cos(\varphi-\varphi_{0})+\cos\theta_{0}\cos\theta],
\end{eqnarray}
where $(\theta_{0},\varphi_{0})$ and $(\theta,\varphi)$ are the
directions of $\textbf{Q}$ and $\textbf{k}^{'}$ in the spherical
coordinate respectively. As a constant phase angle, $\varphi_{0}$
can be eliminated by choosing a special spherical coordinate in
which the direction of $\textbf{Q}$ represents as
$(\theta_{0},\varphi_{0}=0)$.
Only $\theta_{0}$ as a parameter determines the direction of
$\textbf{Q}$.

Substituting the expansion Eq.(1) into Eq.(4) and Eq.(10), one
gets a set of coupled equations for the quantities
$\Delta_{l}^{m_{j}}(k,Q,\theta_{0})$
\begin{eqnarray}
\Delta_{l}^{m_{j}}(k,Q,\theta_{0})=\frac{-1}{\pi}\int_{0}^{\infty}dk^{'}k^{'2}\sum_{l^{'}=0,2}\imath^{l^{'}-l}
V^{l^{'}l}_{\lambda}(k^{'},k)\sum_{l^{''}\mu}\Delta_{l^{''}}^{\mu}(k^{'},Q,\theta_{0})\nonumber\\
\times\int
d\Omega_{\textbf{k}^{'}}Tr[G_{l^{'}}^{m_{j}*}(\textbf{\^{k}}^{'})G_{l^{''}}^{\mu}(\textbf{\^{k}}^{'})]
\frac{1-f(E_{\textbf{k}^{'}}^{+})-f(E_{\textbf{k}^{'}}^{-})}{\sqrt{\xi_{\textbf{k}^{'}}^{2}+D^{2}(\textbf{k}^{'})}}
\end{eqnarray}
with
\begin{eqnarray}
D^{2}(\textbf{k})&&=\frac{1}{2}Tr(\Delta\Delta^{\dag})\nonumber\\
&&=\sum_{ll^{'}=0,2}\sum_{m_{j}m_{j^{'}}}\Delta_{l}^{m_{j}*}(k,Q,\theta_{0})
\Delta_{l^{'}}^{m_{j^{'}}}(k,Q,\theta_{0})Tr[G_{l}^{m_{j}\dag}(\textbf{\^{k}})G_{l^{'}}^{m_{j^{'}}}(\textbf{\^{k}})],\nonumber\\
\end{eqnarray}
where
\begin{eqnarray}
V^{l^{'}l}_{\lambda}(k^{'},k)\equiv<k^{'}\mid
V^{l^{'}l}_{\lambda}\mid k>=\int_{0}^{\infty}r^{2}dr
j_{l^{'}}(k^{'}r)V^{l^{'}l}_{\lambda}(r)j_{l}(k r)
\end{eqnarray}
are the matrix elements of the NN interaction in different partial
wave ($\lambda=T,S,l,l^{'}$) channels. In the present calculation,
$\lambda$ corresponds to the coupled $^{3}SD_{1}$ channel.

In Ref.\cite{min}, we have proposed an axi-symmetric
$D^{2}(\textbf{k})$ solution with an axi-symmetric deformation of
the neutron and proton Fermi spheres. The axi-symmetric
$D^{2}(\textbf{k})$ corresponds to the $m_{j}=0$ gap components of
$\Delta_{l}^{m_{j}}(k)$ only. Moreover, the relations
$\Delta_{l}^{m_{j}*}(k,Q,\theta_{0})=-(-1)^{m_{j}}\Delta_{l}^{-m_{j}}(k,Q,\theta_{0})$
from Eq.(3) yield,
\begin{eqnarray}
\Delta_{0}^{0}(k,Q,\theta_{0})&&=\imath\delta_{0}(k,Q,\theta_{0}),\nonumber\\
\Delta_{2}^{0}(k,Q,\theta_{0})&&=\imath\delta_{2}(k,Q,\theta_{0}).
\end{eqnarray}
Then we can write the axisymmetric $D^{2}(\textbf{k})$ as
\begin{eqnarray}
D^{2}(\textbf{k})&&\rightarrow D^{2}(k,\theta,Q,\theta_{0})=\frac{1}{2}\emph{A}(\theta)[\delta_{0}(k,Q,\theta_{0})]^{2}\nonumber\\
&&-\emph{B}(\theta)\delta_{0}(k,Q,\theta_{0})\delta_{2}(k,Q,\theta_{0})+\frac{1}{2}\emph{C}(\theta)[\delta_{2}(k,Q,\theta_{0})]^{2},
\end{eqnarray}
where
\begin{eqnarray}
\emph{A}(\theta)&&=Tr[G_{0}^{0\dag}(\textbf{\^{k}}^{'})G_{0}^{0}(\textbf{\^{k}}^{'})]=\frac{1}{4\pi},\nonumber\\
\emph{B}(\theta)&&=-Tr[G_{0}^{0\dag}(\textbf{\^{k}}^{'})G_{2}^{0}(\textbf{\^{k}}^{'})]\nonumber\\
&&=-Tr[G_{2}^{0\dag}(\textbf{\^{k}}^{'})G_{0}^{0}(\textbf{\^{k}}^{'})]=\frac{\sqrt{2}}{8\pi}(3\cos^{2}\theta-1),\nonumber\\
\emph{C}(\theta)&&=Tr[G_{2}^{0\dag}(\textbf{\^{k}}^{'})G_{2}^{0}(\textbf{\^{k}}^{'})]=\frac{1}{8\pi}(3\cos^{2}\theta+1).
\end{eqnarray}
The $\theta$ dependent $D^{2}(k,\theta,Q,\theta_{0})$ is
independent of $\varphi$ and maintains the rotational symmetry
[denoted as O(2) symmetry] along the axis $(\theta=0,\varphi=0)$
(the symmetry-axis of ADG), which is also the O(3) symmetry
breaking axis. The pair momentum $2\textbf{Q}$ breaks both
rotational and translational symmetries and the symmetry breaking
axis is $(\theta=\theta_{0},\varphi=0)$. The angle between the two
symmetry breaking axes is $\theta_{0}$ and the two axes are
parallel/perpendicular to each other when
$\theta_{0}=0/\theta_{0}=\frac{\pi}{2}$.

Introducing the normalization
\begin{eqnarray}
\Delta_{s}(k,Q,\theta_{0})=\sqrt{\frac{1}{8\pi}}\delta_{0}(k,Q,\theta_{0}),
\Delta_{d}(k,Q,\theta_{0})=-\sqrt{\frac{1}{8\pi}}\delta_{2}(k,Q,\theta_{0}),
\end{eqnarray}
one gets the $m_{j}=0$ components of the gap equations from
Eq.(12) with finite Cooper pair momentum for the FFLO-ADG state
\begin{eqnarray}
\left(
\begin{array}{l}
\Delta_{s} \\
\Delta_{d}
\end{array}
\right)(k,Q,\theta_{0})=\frac{-1}{\pi}\int dk^{'}k^{'2}\left(
\begin{array}{ll}
V^{00} & V^{02}\\
V^{20} & V^{22}
\end{array}
\right)(k,k^{'})\nonumber\\
\times\int
d\Omega_{\textbf{k}^{'}}\frac{1-f(E_{\textbf{k}^{'}}^{+})-f(E_{\textbf{k}^{'}}^{-})}{\sqrt{\xi_{\textbf{k}^{'}}^{2}+D^{2}(k^{'},\theta,Q,\theta_{0})}}
\left(\begin{array}{ll}
\emph{A}(\theta) & \emph{B}(\theta)\\
\emph{B}(\theta) & \emph{C}(\theta)
\end{array}
\right) \left(
\begin{array}{l}
\Delta_{s} \\
\Delta_{d}
\end{array}\right)(k^{'},Q,\theta_{0}),
\end{eqnarray}
where $V^{00}$, $V^{02}$, $V^{20}$, $V^{22}$ are defined in
Eq.(14) with $l,l^{'}=0,2$ and the quasiparticle spectrum is
\begin{eqnarray}
 E^{\pm}_\textbf{k}&&=\sqrt{\xi_{\textbf{k}}^{2}+D^{2}(k,\theta,Q,\theta_{0})}\nonumber\\
 &&\pm[\delta\mu-\frac{k^{'}Q}{2m}(\sin\theta_{0}\sin\theta\cos\varphi+\cos\theta_{0}\cos\theta)].
\end{eqnarray}
The axi-symmetric $D^{2}(k,\theta,Q,\theta_{0})$ is given by,
\begin{eqnarray}
D^{2}(k,\theta,Q,\theta_{0})=\Delta_{s}^{2}(k,Q,\theta_{0})
+\Delta_{d}^{2}(k,Q,\theta_{0})[\frac{3\cos^{2}\theta+1}{2}]\nonumber\\
+\sqrt{2}\Delta_{s}(k,Q,\theta_{0})\Delta_{d}(k,Q,\theta_{0})[3\cos^{2}\theta-1].
\end{eqnarray}
The two coupled components $\Delta_{s}$ and $\Delta_{d}$ represent
the $^{3}S_{1}$ and $^{3}D_{1}$ channel gaps, respectively. At
variance with the discussion in Ref.\cite{ffn}, the orientation of
the Cooper pair momentum may affect the quasiparticle excitation
and the system becomes more asymmetric if the two symmetry
breaking axes are not parallel to each other.


Following Eq.(5), we can get the neutron and proton densities,
\begin{eqnarray}
\rho^{(p/n)}=\sum_{\textbf{k},\sigma}n_{\sigma}^{(p/n)}(\textbf{k}),
\end{eqnarray}
with their distributions
\begin{eqnarray}
n_{\sigma}^{(p/n)}(\textbf{k})=\{\frac{1}{2}(1+\frac{\xi_{\textbf{k}}}{\sqrt{\xi_{\textbf{k}}^{2}+D^{2}(k,\theta,Q,\theta_{0})}})f(E_{\textbf{k}}^{\pm})\nonumber\\
+\frac{1}{2}(1-\frac{\xi_{\textbf{k}}}{\sqrt{\xi_{\textbf{k}}^{2}+D^{2}(k,\theta,Q,\theta_{0})}})[1-f(E_{\textbf{k}}^{\mp})]
\}.
\end{eqnarray}
Summation over frequencies in Eq.(6) leads to the density matrix
of the particles in the condensate,
\begin{eqnarray}
\nu_{\sigma_{1},\sigma_{2}}(\textbf{k},Q,\theta_{0})=\frac{\Delta_{\sigma_{1},\sigma_{2}}(\textbf{k},Q,\theta_{0})}
{2\sqrt{\xi_{\textbf{k}}^{2}+D^{2}(k,\theta,Q,\theta_{0})}}[1-f(E_{\textbf{k}}^{+})-f(E_{\textbf{k}}^{-})].
\end{eqnarray}
For asymmetric nuclear matter, the coupled equations (19) and (22)
should be solved selfconsistently with the expressions (20) and
(21) for the FFLO-ADG state.
 In our calculation, the total Cooper pair momentum $2Q$ and the angle $\theta_{0}$ are treated as
variational parameters to be determined from the ground state
energy of the system.

\subsection{Thermodynamics}
For asymmetric nuclear matter at a fixed finite temperature and
given neutron and proton densities, the total density
$\rho=(\rho^{^{(n)}}+\rho^{^{(p)}})$ and the isospin asymmetry
$\alpha=(\rho^{^{(n)}}-\rho^{^{(p)}})/\rho$ are fixed. The
thermodynamic quantity describing the system is the free energy
defined as
\begin{eqnarray}
\emph{F}|_{\rho,\beta}=\emph{U}-T\emph{S},
\end{eqnarray}
where $\emph{U}$ is the internal energy and $\emph{S}$ is the
entropy. A thermodynamically stable state minimizes the difference
between the free energies of the superconducting and normal
states, $\delta\emph{f}=\emph{F}_{s}-\emph{F}_{n}$. In the
mean-field approximation, the entropy of the superfluid state is
\begin{eqnarray}
\emph{S}_{s}=-2k_{B}\sum_{\textbf{k}}&&\{f(E_{\textbf{k}}^{+})\ln f(E_{\textbf{k}}^{+})+\bar{f}(E_{\textbf{k}}^{+})\ln \bar{f}(E_{\textbf{k}}^{+})\nonumber\\
&&+f(E_{\textbf{k}}^{-})\ln
f(E_{\textbf{k}}^{-})+\bar{f}(E_{\textbf{k}}^{-})\ln
\bar{f}(E_{\textbf{k}}^{-}) \},
\end{eqnarray}
where $\bar{f}(E_{\textbf{k}}^{\pm})=1-f(E_{\textbf{k}}^{\pm})$
and the summation is over the momentum states in quasiparticle
approximation. Taking the limit $\Delta\rightarrow0$, we get the
entropy $\emph{S}_{n}$ in the normal state. The mean-field
internal energy of the superfluid state reads
\begin{eqnarray}
\emph{U}&&=\sum_{\sigma\textbf{k}}[\varepsilon_{\textbf{k}}^{(n)}n_{\sigma}^{(n)}(\textbf{k})+\varepsilon_{\textbf{k}}^{(p)}n_{\sigma}^{(p)}(\textbf{k})]\nonumber\\
&&+\sum_{\textbf{k},\textbf{k}^{'}}\sum_{\sigma_{1},\sigma_{2},\sigma_{1}^{'},\sigma_{2}^{'}}
<\textbf{k}\sigma_{1},-\textbf{k}\sigma_{2}\mid V\mid\textbf{k}^{'}\sigma_{1}^{'},-\textbf{k}^{'}\sigma_{2}^{'}>\nonumber\\
&&\times\nu^{\dag}_{\sigma_{1},\sigma_{2}}(\textbf{k},Q,\theta_{0})
\nu_{\sigma_{1}^{'},\sigma_{2}^{'}}(\textbf{k}^{'},Q,\theta_{0}).
\end{eqnarray}
The second term in Eq.(27) includes the BCS mean-field interaction
among the particles in the condensate and can be eliminated in
terms of the gap equation (10). Finally, the internal energy is
written as
\begin{eqnarray}
\emph{U}&&=\sum_{\sigma\textbf{k}}[\varepsilon_{\textbf{k}}^{(n)}n_{\sigma}^{(n)}(\textbf{k})+\varepsilon_{\textbf{k}}^{(p)}n_{\sigma}^{(p)}(\textbf{k})]\nonumber\\
&&-\sum_{\textbf{k}}\frac{D^{2}(k,\theta,Q,\theta_{0})}{\sqrt{\xi_{\textbf{k}}^{2}+D^{2}(k,\theta,Q,\theta_{0})}}[1-f(E_{\textbf{k}}^{+})-f(E_{\textbf{k}}^{-})].
\end{eqnarray}
The first term in Eqs.(27) and (28) contains the kinetic energy of
the quasiparticles which is a functional of the pairing gap. In
the normal state, it can be reduced to the kinetic energy of the
neutrons and protons. Noting that the introduction of the Cooper
pair momentum $2\textbf{Q}$ can enhance both the pairing energy
and the kinetic energy. The competition between these two
mechanisms can adjust the value and the direction of $\textbf{Q}$.


\section{Numerical Results}
The nuclear FFLO states with ADG are studied numerically using the
Argonne $V_{18}$ potential. In the present paper, we focus on the
effects due to the finite momentum of the Cooper pairs with
axi-symmetric ADG and investigate the favored angle between the
directions of the Cooper pair momentum and the symmetry-axis of
ADG. Several assumptions are adopted to simplify the calculations.
First, we adopt the free single-particle (s.p.) energy spectrum,
which may affect the level density of the state around Fermi
surface. Second, the pairing interaction is approximated by the
bare interaction, i.e., ignoring the screening effects of the
pairing interaction. These two approximations may affect the
absolute magnitude of the pairing gap. Moreover, we only study the
coupled $^{3}SD_{1}$ channel which dominates the pairing force at
the considered density $\rho_{0}=0.17fm^{-3}$. At this density,
the previous studies \cite{zuo,tt,jme} show that the n-p pairing
gap is about $12$ MeV with the free s.p. spectrum and about $5$
MeV with the Brueckner-Hartree-Fock (BHF) s.p. spectrum, which
seem difficult to reconcile with the empirical information
available from finite nuclei. \cite{fnc}. However, the n-p pairing
gap is reduced to less than $0.5$ MeV at density of
$\rho_{0}\approx0.17fm^{-3}$ when the energy dependence of the
single-particle self-energy (within the BHF approximation) is
taken into account in the gap equation \cite{img}. In this paper,
we adopt the free s.p. spectrum. In order to remove the dependence
on the absolute scale of the gap, we present the results
normalized to the pairing gap in the symmetric matter at zero
total momentum of the pairs. The computations are carried out at
the temperature $\beta^{-1}= 0.5$ MeV at which the effect of the
angle dependence of the pairing gap becomes important\cite{min}.
Both the values of $\theta_{0}$ and $Q$ for the local stable state
are determined by minimizing the free energy.


\begin{figure}
\includegraphics[scale=0.7]{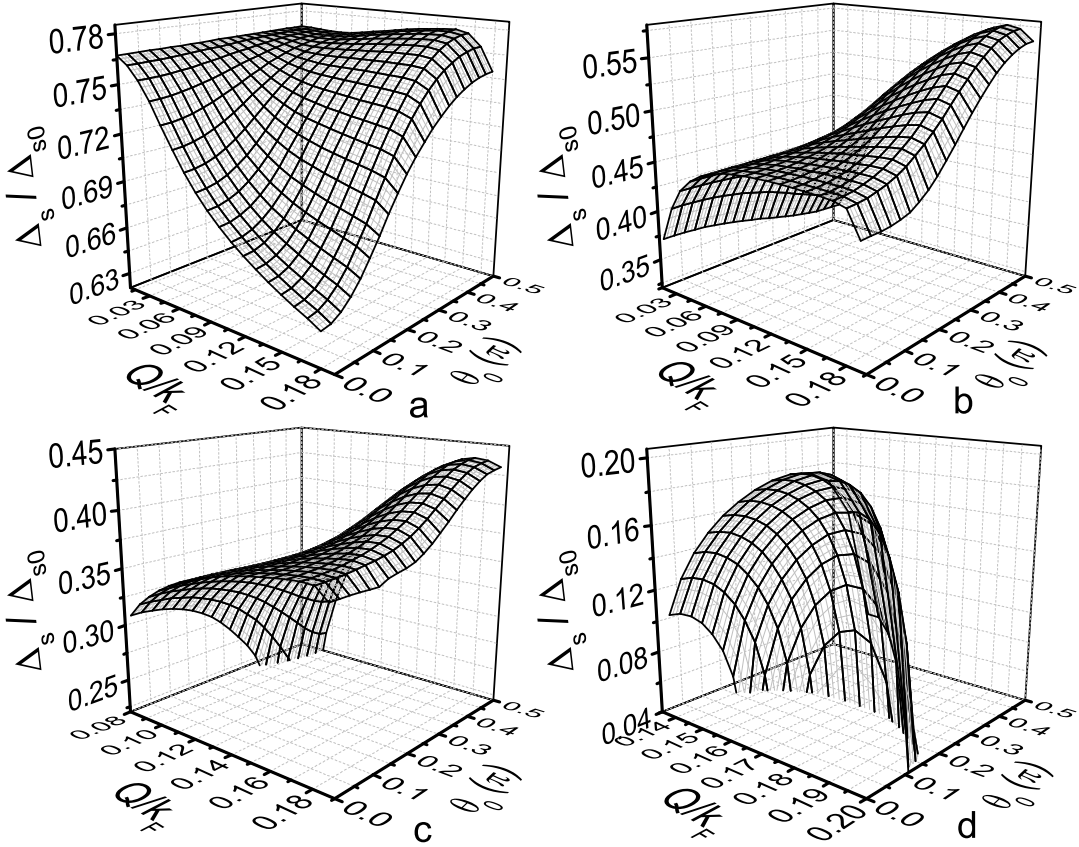} \caption{The value of $\Delta_{s}(k_{F})$
vs $\theta_{0}$ (the angle between the direction of the Cooper
pair momentum and the symmetry-axis of ADG) and $Q$ (the half
value of the Cooper pair momentum) in units of the Fermi momentum
$k_{F}$. The pairing gap is normalized to its value in the
symmetric and zero-total-momentum case $\Delta_{s0}$ at fixed
density $\rho=0.17fm^{-3}$ and temperature $\beta^{-1}= 0.5$ MeV.
Figs.(a), (b), (c), (d) are related to the select isospin
asymmetry $\alpha=$ $0.15$, $0.25$, $0.30$, $0.40$, respectively.
} \label{gap0}
\end{figure}

Fig.1 shows the pairing gap $\Delta_{s}(k_{F})$ in the $^{3}S_{1}$
partial-wave channel as a function of $\theta_{0}$
and $Q$ (in units of the Fermi momentum $k_{F}$).
The isospin asymmetries $\alpha$ are set to be $0.15$, $0.25$,
$0.30$, $0.40$ in Fig.1a, Fig.1b, Fig.1c, Fig.1d, respectively.
The gap is normalized to its value
$\Delta_{s0}=\Delta_{s}(k_{F})[\alpha=0,Q=0]$ with
zero-total-momentum in symmetric nuclear matter. For the small
asymmetry $\alpha=0.15$ in Fig.1a, $\Delta_{s}(k_{F})$ decreases
monotonically as a function of $Q$ at $\theta_{0}=0$. However, for
$0.4\pi\leq\theta_{0}\leq0.5\pi$ $\Delta_{s}(k_{F})$ takes its
maximal value at $Q\neq0$, indicating that the FFLO state can
exist in the domain of $\theta_{0}$ around the direction of the
Cooper pair momentum perpendicular to the symmetry-axis of ADG.
At the two moderate isospin asymmetries $\alpha=$ $0.25$ and
$0.30$ in Figs. 1b and 1c, the maxima of $\Delta_{s}(k_{F})$ are
locateed at $Q\neq0$ for any $\theta_{0}$ of
$0\leq\theta_{0}\leq\frac{\pi}{2}$, indicating a FFLO state for
any orientation of the Cooper pair momentum. Fig.1d shows that
$\Delta_{s}(k_{F})$ take its the maximum at $Q\neq0$ for
$0\leq\theta_{0}\leq0.28\pi$, i.e., the FFLO state exists in a
narrow region of $\theta_{0}$ near the direction of the Cooper
pair momentum parallel to the symmetry-axis of ADG. It is also
shown in Fig.1d that the pairing gap can only exist with nonzero
$Q$, which implies only the FFLO state can survive for
sufficiently large asymmetry (for example, $\alpha=$ $0.40$).



\begin{figure}
\includegraphics[scale=0.7]{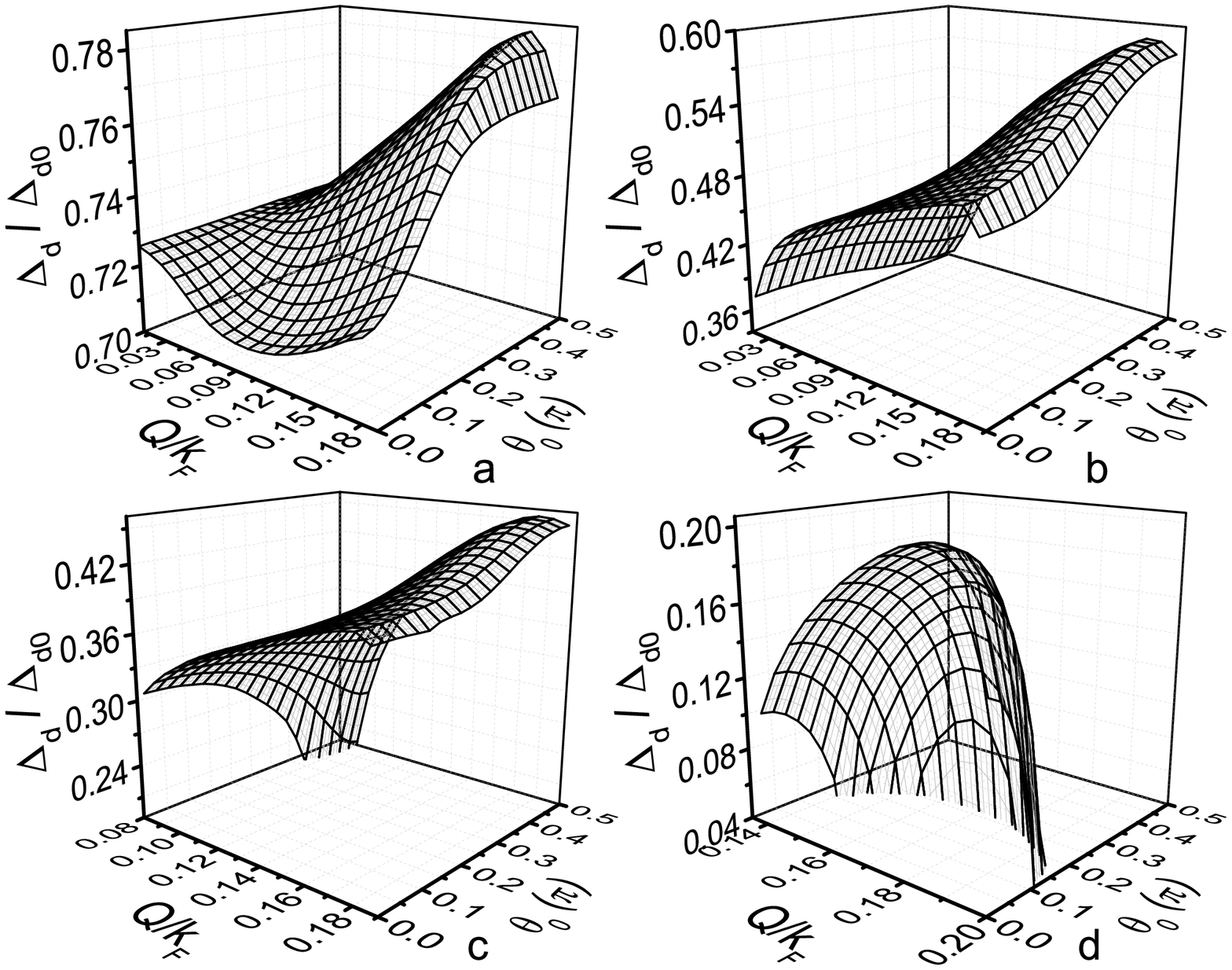} \caption{The value of $\Delta_{d}(k_{F})$
vs $\theta_{0}$ (the angle between the direction of the Cooper
pair momentum and the symmetry-axis of ADG) and $Q$ (the half
value of the Cooper pair momentum) in units of the Fermi momentum
$k_{F}$. The pairing gap is normalized to its value in the
symmetric and zero-total-momentum case $\Delta_{d0}$ at fixed
density $\rho=0.17fm^{-3}$ and temperature $\beta^{-1}= 0.5$ MeV.
Figs.(a), (b), (c), (d) are related to the select isospin
asymmetry $\alpha=$ $0.15$, $0.25$, $0.30$, $0.40$, respectively.}
\label{gap2}
\end{figure}

In the $^{3}D_{1}$ partial-wave channel, the pairing gap
$\Delta_{d}(k_{F})$ also plays an important role in the
condensate. Therefore, we show $\Delta_{d}(k_{F})$ as a function
of $\theta_{0}$ and $Q$ (in units of the Fermi momentum $k_{F}$)
in Fig.2, where the parameters are set to be the same as those in
Fig.1 and the gap value is normalized to its value
$\Delta_{d0}=\Delta_{d}(k_{F})[\alpha=0,Q=0]$ with
zero-total-momentum in symmetric nuclear matter.
The shapes of the $\Delta_{d}(k_{F})$ surfaces closely resemble
those of the pairing gap $\Delta_{s}(k_{F})$ in Fig.1 except for
Fig.2a. In Fig.2a, there exists a minimum of $\Delta_{d}(k_{F})$
at a nonzero $Q$ for $\theta_{0}=0$, which may indicate that the
FFLO state can not exist for a small asymmetry when the Cooper
pair momentum is parallel to the symmetry-axis of ADG.

\begin{figure}
\includegraphics[scale=0.7]{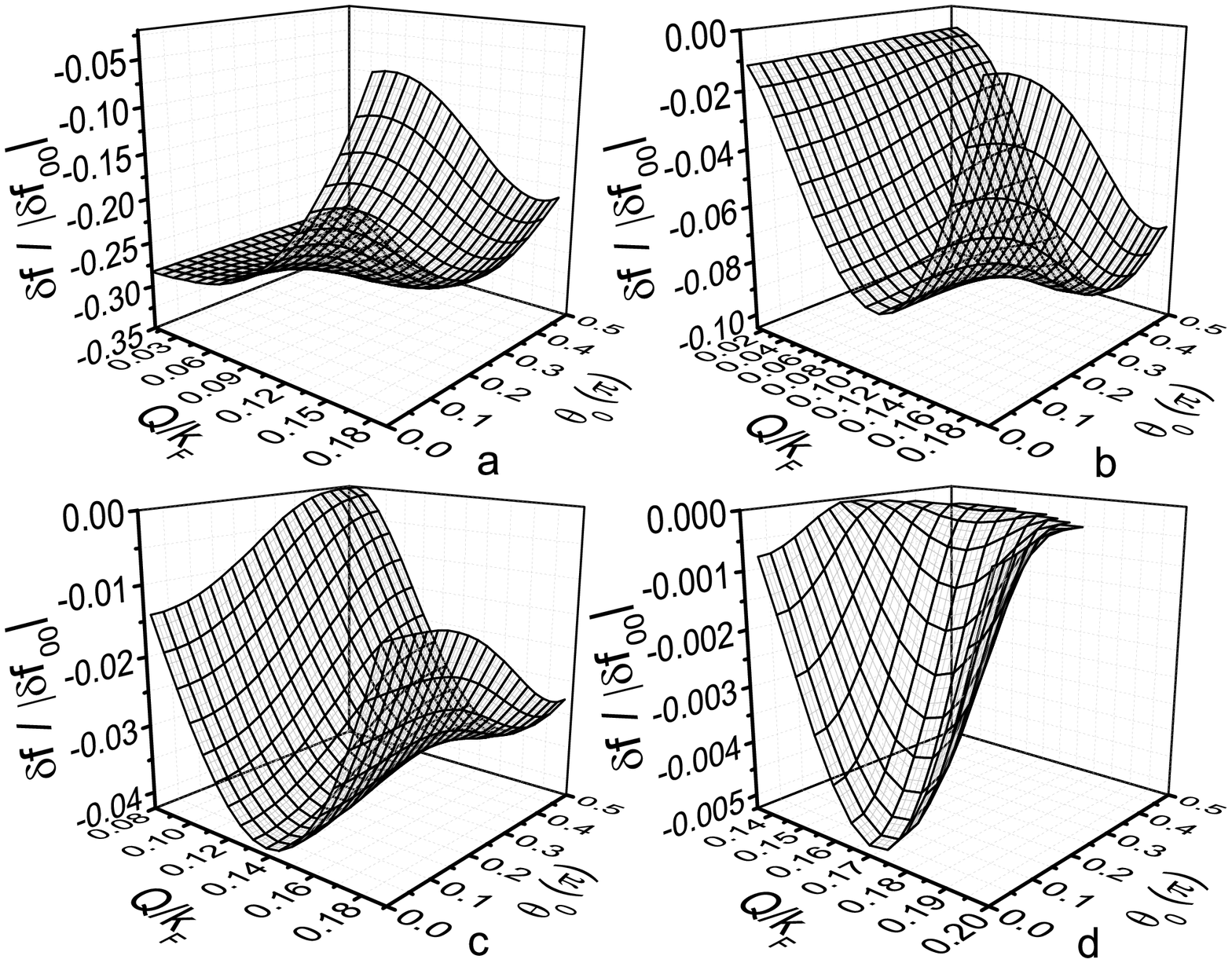} \caption{The difference of the free energy of the normal and superconductiong state $\delta\emph{f}$
as a function of $\theta_{0}$ and $Q$, where $Q$ have been
normalized to the Fermi momentum $k_{F}$ and $\theta_{0}$ is the
angle between the direction of the Cooper pair momentum and the
symmetry-axis of ADG. $\delta\emph{f}$ is normalized to its value
in the symmetric and zero-total-momentum case
$\mid\delta\emph{f}_{00}\mid$ at fixed density $\rho=0.17fm^{-3}$
and temperature $\beta^{-1}= 0.5$ MeV. Figs.(a), (b), (c), (d) are
related to the select isospin asymmetry $\alpha=$ $0.15$, $0.25$,
$0.30$, $0.40$, respectively.} \label{e0}
\end{figure}

In order to find the ground state, we calculate the free-energy
difference $\delta\emph{f}$ between the normal and superconducting
states as a function of $\theta_{0}$ and $Q$. The results are
shown in Fig. 3, where the parameters are also set to the same
values as those in Fig. 1. The free energy difference
$\delta\emph{f}$ is normalized to its value at zero total momentum
in symmetric matter. At $\alpha=0.15$ in Fig.3a, the sole local
minimum of $\delta\emph{f}$ is located
at $(\theta_{0}=\frac{\pi}{2},Q=0.097)$, indicating that the FFLO
state is stable when the direction of the Cooper pair momentum and
the symmetry-axis of ADG are orthogonal for a small $\alpha$.
Below we call this state as FFLO-ADG-Orthogonal state. In Figs.3b,
there exist two local minima of $\delta\emph{f}$ settled at
$(\theta_{0}=\frac{\pi}{2},Q=0.145)$ and $(\theta_{0}=0,Q=0.18)$,
respectively. The second local minimum is related to the case that
the directions of the Cooper pair momentum and the symmetry-axis
of ADG are parallel. We call this state FFLO-ADG-Parallel state
below. For moderate asymmetries (for examples, $\alpha=$ $0.25$,
$0.30$), both the FFLO-ADG-Orthogonal state and the
FFLO-ADG-Parallel state are locally stable. Comparing Fig.3b and
Fig.3c, the FFLO-ADG-Orthogonal state is more stable at smaller
$\alpha$, whereas the FFLO-ADG-Parallel state is more favored for
larger $\alpha$. Fig.3d shows only the FFLO-ADG-Parallel state can
survive at a sufficiently large $\alpha$.

\begin{figure}
\includegraphics[scale=0.7]{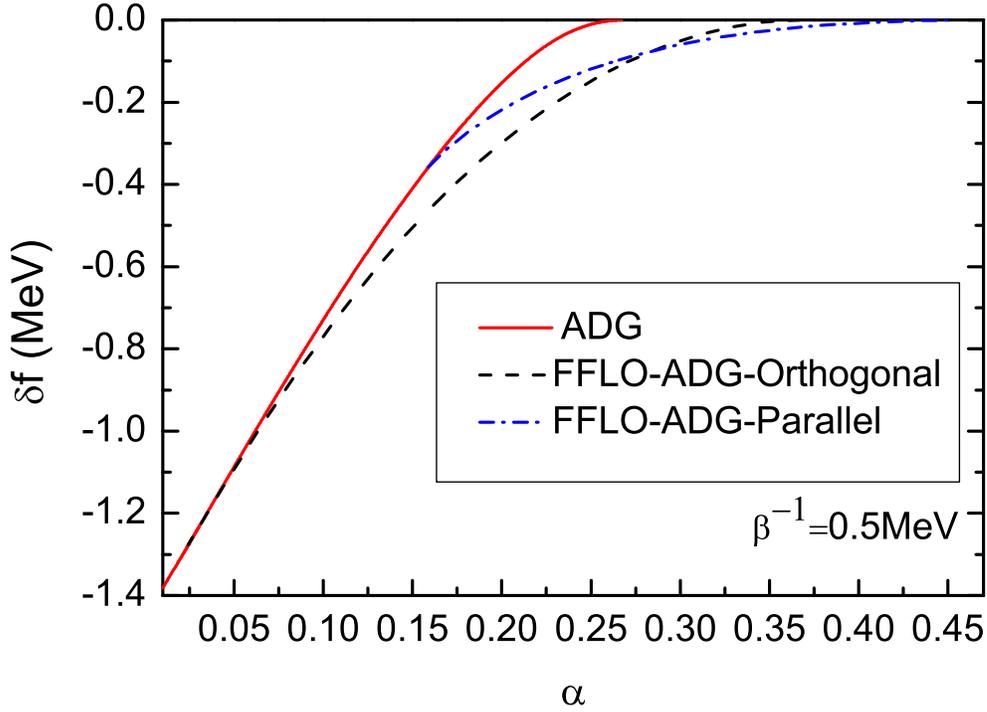} \caption{(Color online).The phase diagram as a function of asymmetry $\alpha$. The red solid, black dashed and the blue dash-dotted line are relate to the ADG, FFLO-ADG-Orthogonal and FFLO-ADG-Parallel state, respectively. }
\label{dife}
\end{figure}

We can find from Fig.3 that the minimum of the free-energy
difference $\delta\emph{f}$ is locared either at $\theta_{0}=0$ or
$\theta_{0}=\frac{\pi}{2}$. In fact, we have calculated
$\delta\emph{f}$ by varying $\theta_{0}$ and $Q$ for
different asymmetries $\alpha$ ($0.01\leq\alpha\leq0.47$). The
results show the minimum of $\delta\emph{f}$ is related to either
$\theta_{0}=\frac{\pi}{2}$ or $\theta_{0}=0$, indicating that only
the FFLO-ADG-Parallel state and the FFLO-ADG-Orthogonal state are
locally stable. Fig.4 displays the calculated values of the local
minimum of $\delta\emph{f}$ vs. the asymmetry $\alpha$. For
comparison, the result for the ADG state corresponding to $Q=0$ is
also exhibited in Fig.4 (the red solid line). The black dashed
line is related to $(\theta_{0}=\frac{\pi}{2},Q\neq0)$, i.e., the
FFLO-ADG-Orthogonal state. And the blue dash-dotted line
corresponds to $(\theta_{0}=0,Q\neq0)$, i.e., the
FFLO-ADG-Parallel state. For small isospin asymmetries $\alpha$
($0.025<\alpha<0.16$), there exists only one kind of locally
stable FFLO state, i.e., the FFLO-ADG-Orthogonal state. With
increasing $\alpha$, the FFLO-ADG-Parallel state emerges above
$\alpha=0.16$,
but it is not the favored state for $\alpha<0.286$. The
FFLO-ADG-Parallel state becomes more stable than the
FFLO-ADG-Orthogonal state when $\alpha>0.286$. At $\alpha=0.37$,
the FFLO-ADG-Orthogonal state vanishes, whereas the
FFLO-ADG-Parallel state can exist up to $\alpha=0.47$.

Since the contribution to superfluidity from the Cooper pairs
around the average Fermi surface is dominant, for n-p pair
correlation, a small separation of the neutron and proton Fermi
surfaces $\delta\mu$ may suppress the superfluidity strongly. In
the ADG configuration, the neutron Fermi sphere possesses an
oblate deformation along the symmetry-axis of ADG, whereas the
proton Fermi sphere has a prolate deformation. These two different
kinds of deformation enhance the correlation between neutrons and
protons near their average Fermi surface, i.e., the phase space
near $(\theta=0,\varphi=0)$ and $(\theta=\pi,\varphi=0)$. In the
FFLO configuration, the shift of the neutron and proton Fermi
spheres with respect to each other may enhance the overlap between
the two Fermi surfaces. However, the influence of the Cooper pair
momentum turns out to be complicated when considering the angle
dependence of the pairing gap. For example, in the weakly
isospin-asymmetric case, the difference between the neutron and
proton Fermi surfaces $\delta\mu$ is small, and the deformation of
the Fermi spheres in the ADG configuration is sufficient to
compensate the effect due to this difference $\delta\mu$. As a
consequence, a shift of the two Fermi spheres with respect to each
other along the symmetry-axis of ADG ($\theta_{0}=0$) may even
reduce the overlap of the phase-space for pairing near the average
Fermi surface. Therefore, the FFLO-ADG-Parallel state could not
exist for small $\alpha$. Nevertheless, if the shift between the
two Fermi spheres is perpendicular to the symmetry-axis of ADG,
the pairing could be enhanced in three areas in the phase space,
i.e., the area near $(\theta=0,\varphi=0)$,
$(\theta=\pi,\varphi=0)$ and $(\theta=\frac{\pi}{2},\varphi=0)$.
Therefore, only the FFLO-ADG-Orthogonal state is stable for small
asymmetry $\alpha$.

When the asymmetry $\alpha$ gets large, the splitting of the
neutron and proton Fermi surfaces becomes so large that its effect
can not be completely compensated by the effect of the deformation
of the two Fermi spheres in the ADG configuration. The Cooper pair
momentum along the direction of $\theta_{0}=0$ is expected to
reduce the remained splitting partially. Consequently, the
FFLO-ADG-Parallel state emerges and becomes local stable. However,
the splitting is not large enough to destroy the mechanism of the
FFLO-ADG-Orthogonal state totally. Hence, both the
FFLO-ADG-Parallel state and FFLO-ADG-Orthogonal state are local
stable for a moderate $\alpha$. At a sufficiently large asymmetry,
the neutron and proton Fermi surfaces are split so much that the
deformation of the two Fermi spheres in the ADG configuration can
no longer guarantee the pairing around $(\theta=0,\varphi=0)$ and
$(\theta=\pi,\varphi=0)$ near the average Fermi surface. Under
this condition, the neutron and proton Fermi surfaces are much
closer along the direction $(\theta=0,\varphi=0)$ than that along
the direction $(\theta=\frac{\pi}{2},\varphi=0)$ in the ADG
configuration. A shift between the two Fermi spheres along the
direction $(\theta=0,\varphi=0)$ is much easier to promote the
phase-space overlap for the pairing than that along the direction
$(\theta=\frac{\pi}{2},\varphi=0)$, i.e., the FFLO-ADG-Parallel
state is more favored than the FFLO-ADG-Orthogonal state for a
sufficiently large $\alpha$.

\begin{figure}
\includegraphics[scale=0.7]{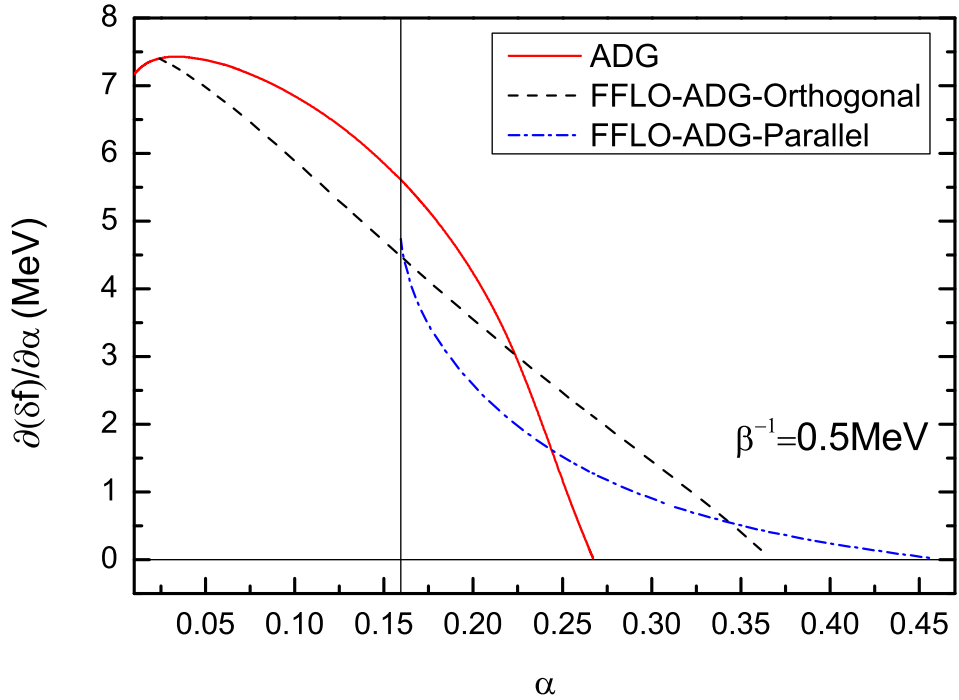} \caption{(Color online). $\frac{\partial\delta\emph{f}}{\partial\alpha}$ vs $\alpha$. The red solid, black dashed and the blue dash-dotted line are relate to the ADG, FFLO-ADG-Orthogonal and FFLO-ADG-Parallel state, respectively. }
\label{ddife}
\end{figure}

In Fig.4, the phase transition from the FFLO-ADG-Orthogonal state
to the FFLO-ADG-Parallel state is of the first order. In order to
investigate in detail the phase transitions in Fig.4, we calculate
$\frac{\partial\delta\emph{f}}{\partial\alpha}$ as function of
$\alpha$ and the results are shown in Fig.5. At $\alpha=0.025$,
the transition from the ADG state to the FFLO-ADG-Orthogonal state
is of the second order. But at $\alpha=0.16$,
$\frac{\partial\delta\emph{f}}{\partial\alpha}\mid_{ADG}\neq\frac{\partial\delta\emph{f}}{\partial\alpha}\mid_{FFLO-ADG-Parallel}$,
which indicates a first-order transition from the ADG state to the
FFLO-ADG-Parallel state. In Fig.4, the three curves of ADG,
FFLO-ADG-Orthogonal and FFLO-ADG-Parallel tend to zero gently near
the transition points, indicating that the three transitions from
the superconducting states to the normal state are of second
order. Fig.5 shows
$\frac{\partial\delta\emph{f}}{\partial\alpha}\rightarrow0$ near
the three phase transition points, which are agreement with the
results in Fig.4.

Before summary, we briefly discuss the difference between the ADG
state and the DFS state. As we know, the quasiparticle spectra in
both ADG and DFS states are angle-dependent. In the DFS state, the
angle dependence of the quasiparticle spectrum results from the
angle dependence of the deformed neutron and proton Fermi surfaces
\cite{dfs,dfsn,dfsa}. However, in the ADG state the pairing gap is
angle-dependent due to the noncentral $^{3}SD_{1}$ channel pairing
force \cite{min}. And the angle-dependent pairing gap leads to the
angle dependence of the quasiparticle spectrum. These two
mechanisms resulting in the angle-dependent quasiparticle spectrum
are different.

\section{Summary and Outlook}
The fermionic condensation in asymmetric nuclear matter leads to
superconducting states with anisotropic Fermi spheres (such as the
FFLO and ADG\cite{min} states). In the ADG state, the angle
dependence of the pairing gap may result in the deformation of the
neutron and proton Fermi spheres. Moreover, the FFLO state
corresponds to a shift of the two Fermi spheres with respect to
each other. In this paper, we consider these two mechanisms
together and investigate the FFLO-ADG state with arbitrary angle
between the direction of the Cooper pair momentum and the
symmetry-axis of ADG. Two kinds of local stable states are found.
One corresponds to the direction of the pair momentum
perpendicular to the symmetry-axis of ADG (the FFLO-ADG-Orthogonal
state), the other is related to the direction of the pair momentum
parallel to the symmetry-axis of ADG (the FFLO-ADG-Parallel
state). It is shown the FFLO-ADG-Orthogonal state possesses the
lowest free energy at small isospin asymmetries, whereas the
FFLO-ADG-Parallel state is favored for large asymmetries. The
transitions from both the ADG and FFLO-ADG-Orthogonal states to
the FFLO-ADG-Parallel state are of the first order. On the
contrary, the transition from the ADG state to the
FFLO-ADG-Orthogonal state is of the second order. Moreover, the
transitions from the three superconducting states (ADG,
FFLO-ADG-Orthogonal and FFLO-ADG-parallel states) to the normal
state are of the second order.

In the previous studies such as Ref.\cite{ffn,dfsn,Sdd2,zuo}, the
effect of the angle dependence of the pairing gap is abandoned,
i.e., supposing the neutron and proton spheres as isotropic ones.
Thus, the FFLO states are degenerate for arbitrary directions of
the Cooper pair momentum. In fact, for the n-p pairing in
asymmetric nuclear matter, the angle dependence of the pairing gap
can compensate the effect due to the splitting of the neutron and
proton Fermi surfaces\cite{min}. In the axi-symmetric ADG
configuration, one particular direction (the symmetry-axis of ADG)
is selected. In this case, the FFLO state with angle-dependent gap
is nondegenerate for the orientations of the Cooper pair momentum.
Only two orientations are shown to be local stable, corresponding
to the FFLO-ADG-Orthogonal and FFLO-ADG-Parallel states,
respectively. As pointed out in Ref.\cite{min} that the ADG
configuration is related to the deformation of the neutron and
proton Fermi spheres. The FFLO-ADG-Parallel/Orthogonal state may
correspond to more complicated deformation of the neutron and
proton Fermi spheres. In this improved calculation, both the value
of the pairing gap and the domain of $\alpha$ in which the
superconducting state exists become large. Especially for the
temperature $\beta^{-1}= 0.5$ MeV, the ADG state vanishes at
$\alpha=0.267$, whereas the FFLO-ADG-Parallel state vanishes at
$\alpha=0.47$. The combination of the FFLO and ADG states promotes
the onset of n-p pairing for large asymmetry. Here we only
consider the nuclear saturation density $\rho_{0}=0.17fm^{-3}$ at
the low temperature of $\beta^{-1}= 0.5$ MeV. The properties of
FFLO-ADG-Parallel/Orthogonal state are expected to be similar for
high densities at low temperature. At low densities and/or high
temperatures the effect of the angle dependence of the pairing gap
becomes weak, and the difference among the ADG, FFLO-ADG-Parallel
and FFLO-ADG-Orthogonal states may become unobvious.

In ADG state, the rotational symmetry is spontaneously broken [the
O(3) symmetry breaks down to the O(2) symmetry]. Moreover, the
translation and rotational symmetries are both broken in the
FFLO-ADG-Parallel state, however, the O(2) rotational symmetry is
maintained. In the FFLO-ADG-Orthogonal state, the O(2) rotational
symmetry is broken as well. As is well known, the continuous
symmetry breaking leads to collective excitations with vanishing
minimal frequency (Goldstone's modes). The symmetry breaking of
the FFLO-ADG-Parallel/Orthogonal state may imply new complicated
collective bosonic modes in asymmetric nuclear matter (except the
collective motion of the Cooper pairs). We only consider the
simplest FF state ($\Delta(\textbf{r})=\Delta e^{\imath
\textbf{q}\cdot\textbf{r}}$) with ADG in the paper. In fact, the
FFLO state is much more complicated, the structure of the true
ground state remains for the future work.

\section*{Acknowledgments}
{The work is supported by the Special Foundation for  theoretical
physics Research Program of China (No. 11347151), the 973 Program
of China under No. 2013CB834405, the National Natural Science
Foundation of China (No. 11435014, 11175219, 11204323), and the
Knowledge Innovation Project(No. KJCX2-EW-N01) of the Chinese
Academy of Sciences.}

\end{document}